\documentclass[aps,prl,twocolumn,groupedaddress,showpacs]{revtex4}

\usepackage[dvips]{graphicx,color}

\begin{document}
\title{A quantum model of space-time-matter}
\author{Isaac Cohen}
\email[]{icohen@brandeis.edu}
\affiliation{Physics Department, Brandeis University, Waltham, MA
02453}

\date{\today}

\begin{abstract}
We study a quantum mechanics with the usual postulates but in which
the Heisenberg algebra of canonical commutation relations and the
Poincare algebra are replaced by the Lie algebra of the homogeneous
Lorentz group SO(5,1). It arises from the hypothesis that the above
group is the fundamental group of invariance for the laws of
physics. The observables of the theory like position, time,
momentum, energy, angular momentum and others are the generators of
the algebra of the group. Neither position and time observables
commute between them, nor momentum and energy observables. The
algebra of Poincare quantum mechanics is recovered in the limit in
which two parameters, that we physically interpret as the Hubble
constant and the Planck mass, are taken to zero and infinite
respectively. We consider the equations that are satisfied by the
spinor representation of the group.
\end{abstract}\

 \pacs{02.40.Gh, 03.30.+p, 04.60.-m, 11.15.-q, 98.80.Qc}

\maketitle


A satisfactory quantum theory of space-time is lacking. One may
trace the difficulty to obtain such a theory to the early history of
the theory of quantum mechanics as it was developed by Heisenberg,
Schroedinger, Dirac and others. In the study of a physical system,
say a particle in a box, one assumes a classical physical space-time
as its framework and the correspondence principle is then used to
obtain the canonical commutation rules of the theory and to write
the  equation of evolution. But what to assume in the case that the
physical system is the space-time itself?  Moreover, is it possible
to separate clearly space-time from matter in a consistent quantum
treatment of both? If we turn our attention to space-time, it is
worth to remember how Einstein found a  road to that beautiful place
in which the abysm between space and time that existed in Newton's
mechanical view vanished, and space and time became fused into the
Minkowski space-time continuum. One may then, dream of a quantum
path, close to that of Einstein, that lead us to a place where the
conceptual gap between space-time and matter is also absent.

To try to convert such a hope in reality we need, first, to pay
attention, at what is required to define a quantum theory: We have
to select the observables, give the commutation relations between
them, i.e., its algebra, and finally, we have to specify its
dynamics. A set of observables appropriate for the description of
space-time-matter~\cite{wey} should include the position and time
observables as well as the observables commonly associated with
matter, as the energy, momentum and angular momentum.

Next, we expect that if space, time, and matter are conceptually
similar they would be reflected in the properties of the algebra.
Knowing of the connection between symmetry transformations and
observables, we have, then, to face the apparent existence of two
distinct types of motion, to wit, rotations~\cite{note1} and
translations. One might think that the essential characteristic of
rotations, the periodicity, is absent in translations, but in
crystallography  we know how to make compatible translations and
periodicity. Nevertheless, if we  view translations as rotations we
are led to accept their non-commutativity~\cite{note2}. Being the
generators of space-time translations the momentum observables in
quantum mechanics, we have, then, that the different components of
the momentum and the energy don't commute between them. We would be
departing from the Poincare algebra in which these objects commute.
But we may think that the Poincare algebra is only an approximation
to reality. Similarly, if we regard the position and time
observables as rotations we are confronted with the
non-commutativity of the different position components and the time.
Besides, if position observables and momentum observables are now
rotations, the canonical commutation rules between position and
momentum observables will be valid only approximately. Finally, the
generators of space-time rotations are the angular momentum
observables and with the boosts constitute the algebra of the
homogeneous Lorentz group SO(3,1).

The question that emerges is, whether  there is  a rotation algebra
that contains all these fourteen generators and that reproduces at
least at some limit the known properties of these observables. We
found that the smallest appropriate algebra is the algebra of the
group SO(5,1). The assignment of the remaining generator and the
boosts to its corresponding observables will be done later.

The path we were looking for, salvo a mirage, is now visible.
Minkowski space-time is a necessary consequence of the Principle of
Special Relativity, that states that the laws of physics are
invariant under the group of inhomogeneous Lorentz transformations
SO(3,1), i.e., the Poincare group. Lorentz transformations leave
invariant the differential form
\begin{eqnarray}
ds^{2}=-dt^{2}+ \frac{1}{c^{2}}(dx^{2}+dy^{2}+dz^{2}), \label{eq1}
\end{eqnarray}
for the interval ds between two events. The corresponding form for
the group SO(5,1) is a global form that we shall write in a
convenient way and its invariance will constitute the starting point
of the model.

It is the purpose of  this note to study a quantum mechanics with
the usual postulates but in which the Heisenberg algebra of
canonical commutation relations and the Poincare algebra are
replaced by the Lie algebra of the homogeneous Lorentz group
SO(5,1). It arises from the hypothesis that the above group is the
fundamental group of invariance for the laws of physics. The
observables of the theory like position, time, momentum, energy,
angular momentum and others are the generators of the algebra of the
group. Neither position and time observables commute between them,
nor momentum and energy observables. The algebra of Poincare quantum
mechanics is recovered in the limit in which two parameters, that we
physically interpret as the Hubble constant and the Planck mass are
taken to zero and infinite respectively. We consider the equations
that are satisfied by the spinor representation of the group.

We may  realize the group SO(5,1) as maps onto itself of a
hypersurface in $R^{5,1}$
\begin{eqnarray}
-(c t)^{2}+
 x^{2}+y^{2}+z^{2}+(\frac{c u}{H})^{2}+(\frac{\hbar
  v}{m_{pl} c})^{2}=\frac{\hbar}{H m_{pl}}
 ,  \label{hyp}
\end{eqnarray}
with c and  $\hbar$ being the speed of light and Planck's constant
on one hand, and m and H being the Planck mass and the Hubble
constant. The coordinates u and v are dimensionless. Even tough the
coordinates $x,~ y,~ z~ and~ t$ have dimensions of space and time
respectively, they should be taken as the physical space-time
variables.

The generators $L_{ij}$ of the algebra satisfy the commutation
relations
\begin{eqnarray}
 L_{ij}&=&i(x_{i}\partial_{j}-x_{j}\partial_{i}), \label{alg1} \\
 {[L_{ij},L_{kl}]}&=&i ( \eta_{jk}L_{il}-\eta_{ik}L_{jl}-\eta_{jl}L_{ik}
  +\eta_{il}L_{jk})  ,    \label{alg2}
\end{eqnarray}
with $i, j, k, l=0, 1, 2, 3, 5, 6.$ The nonzero elements of the
metric $\eta_{ij}$ are $\eta_{00}=-1$ and
$\eta_{11}=\eta_{22}=\eta_{33}=\eta_{55}=\eta_{66}=1$. Here
$x^{i}\equiv(x^{0},x^{1},x^{2},x^{3},x^{5},x^{6})\equiv(c
t,~x,~y,~z,~\frac{c u}{H},~\frac{\hbar v}{m_{pl}c})$.

The invariance of the form~(\ref{hyp}) may be resumed in the
following two postulates:

\textit{1- The laws of physics are the same for all observers.}

\textit{2- The speed of light c, the planck constant $\hbar$, the}

\textit{Hubble constant H and the Planck mass $m_{pl}$ are the}

\textit{same for all observers.}

We associate with a given observer a set of observables. These
physical observables are identified with the set of hermitian
operators $L_{ij}$ of the algebra of the Lorentz group SO(5,1) as
follows:

1) The physical  position three-vector
$\mathbf{X}=\{X^{1},X^{2},X^{3}\}$ and the physical time $T=X^{0}$
are represented  by the three-vector operator
$\frac{\hbar}{m_{pl}c}\{L^{61},L^{62},L^{63}\}$ and
$\frac{\hbar}{m_{pl}c^{2}}L^{60}$ respectively.

2) The  momentum three-vector $\{P^{1},P^{2},P^{3}\}$ and the energy
$ P^{0}$ are represented by the three-vector operator $
\mathbf{P}=\frac{\hbar H}{c}\{L^{51},L^{52},L^{53}\}$ and $\hbar H
L^{50}$ respectively.

3) The angular momentum three-vector
$\mathbf{J}=J^{23},J^{31},J^{12}$ is represented by the three-vector
operator $ \hbar\{L^{23},L^{31},L^{12}\}$and the  three-vector
operator $\frac{\hbar}{c}\{L^{01},L^{02},L^{03}\}$ represents a
three-vector observable $\{J^{01},J^{02},J^{03}\}$ that we suggest
is the acceleration three-vector~\cite{note3}.

4) Finally, the mass scale S is represented by $\frac{H}{m_{pl}}
J^{56}$.

In natural units, where the speed of light c, and the Planck
constant $\hbar$ are set to 1, the commutation
relations~(\ref{alg2}) read
\begin{eqnarray}
 [P_{\mu},P_{\nu}]&=&i~  H^{2}J_{\mu \nu}, \label{p1} \\
 {[J_{\mu \nu},P_{\rho}]}&=&-i~  \eta_{\mu \rho}
 P_{\nu}+ i~  \eta_{\nu \rho}P_{\mu},
  \label{p2}
\end{eqnarray}
\begin{eqnarray}
 [X_{\mu},X_{\nu}]&=&  \frac{i}{m_{pl}^{2}}J_{\mu \nu},\label{p3}\\
 {[J_{\mu \nu},X_{\rho}]}&=&-i~  \eta_{\mu \rho}
 X_{\nu}+ i~  \eta_{\nu \rho}X_{\mu},
  \label{p4}
\end{eqnarray}
\begin{eqnarray}
 [J_{\mu \nu},J_{\rho \sigma}]=i \eta_{\nu \rho} J_{\mu \sigma}
 -i \eta_{\nu \rho} J_{\nu \sigma}-i\eta_{\nu \sigma} J_{\mu \rho}
 +i \eta_{\mu \sigma} J_{\nu \rho}, \label{p5}
\end{eqnarray}
\begin{eqnarray}
 [X_{\mu},P_{\nu}]&=&i \eta_{\mu\nu} S,  \label{hei} \\
 {[S,X_{\mu}]}&=&\frac{i}{m_{pl}^{2}}P_{\mu}, \label{s1} \\
 {[S,P_{\mu}]}&=&-i H^{2}  X_{\mu},  \label{s2} \\
 {[S,J_{\mu \nu}]}&=&0, \label{s3}
\end{eqnarray}
with $\mu,\nu=0,1,2,3.$

The components of the energy-momentum relations~(\ref{p1}) as well
as the components of the position and time~(\ref{p3}) do not
commute. They represent a departure of the Poincare algebra, but
when the Planck mass $m_{pl}\rightarrow\infty$ and the Hubble
constant $H\rightarrow 0$ their commutation is restored  and the set
of commutation relations~(\ref{p1})-(\ref{p5}) gives the Poincare
algebra. This is the basic reason for the physical interpretation of
the parameters $m_{pl}$ and H, because it is known that the Poincare
algebra should be just an approximation at large scales, i.e., at
the scale of the Hubble constant H, and at short scales, i.e., at
the scale of the Planck mass $m_{pl}$.

The mass scale S in the commutation relations~(\ref{hei})-(\ref{s3})
replaces the identity I in the Heisenberg algebra. The
non-commutativity of the mass scale S as reflected in the
relations~(\ref{hei})-(\ref{s3}) means a departure of that algebra,
only to be restored in the same limit $m_{pl}\rightarrow\infty$ and
$H\rightarrow 0$. In that case, the observable S commutes with all
the generators of the algebra. If we consider an irreducible
representation of the group, it is a multiple to the identity.

On the other hand, the quadratic Casimir operator associated to the
Lie algebra of the Lorentz group SO(5,1) is
\begin{eqnarray}
\frac{1}{m_{pl}^{2}}P^{2} +H^{2}X^{2} +\frac{H^{2}}{m_{pl}^{2}}
\eta_{\mu \nu}J^{\mu \nu}
 + S^{2} &=&\frac{H^{2}}{m_{pl}^{2}}C_{2}, \label{cas1}
\end{eqnarray}
with $P^{2}= \eta_{\mu \nu}P^{\mu} P^{\nu}$ and $X^{2}= \eta_{\mu
\nu}X^{\mu} X^{\nu}$. $C_{2}$ is a multiple of the identity that
depends on the representation of the group.

If we take the limit~$H\rightarrow0$ and~$m_{pl}\rightarrow\infty$
but in such a way that $\frac{H}{m_{pl}}\rightarrow0$, then the
Eq.~(\ref{cas1}) becomes $S^{2}=0$. All the observables commute now,
and then we may conclude saying that it reached the classical
Poincare geometry.

If we take~$H\rightarrow0$ and~$m_{pl}\rightarrow\infty$ but such
that $\frac{H}{m_{pl}}\rightarrow s$, with s  finite, we get instead
\begin{eqnarray}
\eta_{\mu \nu}J^{\mu \nu}
 +\frac{1}{s^{2}} S^{2} &=&C_{2}, \label{cas2}
\end{eqnarray}
It appears possible to choose s such that the mass scale S becomes
the identity I. We may conclude now that we reached the algebra of
Poincare quantum mechanics.

If we only take  the Hubble constant $H\rightarrow0$, leaving the
Planck mass~$m_{pl}$ finite, the Eq.~(\ref{cas1}) becomes
\begin{eqnarray}
P^{2} + m_{pl}^{2} S^{2} &=&0. \label{cas3}
\end{eqnarray}
We may find  the correspondence rules in the same limit. It is
convenient to be at the point with coordinates $t=x=y=z=v=0$, u=1.
In such a limit, the observable $P_{\mu}$ becomes
\begin{eqnarray}
 P_{\mu}= H L_{5\mu}=i H(\frac{u}{H}\partial_{\mu}- H x_{\mu}\partial_{u})
 \rightarrow i \partial_{\mu},  \label{cr1}
\end{eqnarray}
when the Hubble constant $H\rightarrow0$. Similarly , the mass
observable S becomes,
\begin{eqnarray}
 S= H L_{6 5}=i\frac{H}{m_{pl}} (\frac{H v}{m_{pl}}\partial_{u}- \frac{u m_{pl}}{H} \partial_{v})
 \rightarrow -i \partial_{v},  \label{cr2}
\end{eqnarray}
If one assumes that the wave-function in the v-direction is just a
plane wave $\Phi(v)\sim\exp(iv)$, one obtains $S~=~I$ and the
Eq.~(\ref{cas3}) becomes Einstein's energy-momentum mass relation,
where $m_{pl}$ takes the role of invariant mass of the
representation. We may be confident that the physical interpretation
of the observable S as a mass scale makes sense.

On the other hand in the limit in which the Planck mass
$m_{pl}\rightarrow
 \infty$, the observable $X_{\mu}$ becomes
\begin{eqnarray}
 X_{\mu}= \frac{L_{6\mu}}{m_{pl}} = \frac{i}{m_{pl}}(\frac{v}{m_{pl}}\partial_{\mu}- m_{pl} x_{\mu}\partial_{v})
 \rightarrow -ix_{\mu}\partial_{v}.  \label{cr3}
\end{eqnarray}
With the same behavior of the wave-function in the v-direction one
obtains, $X_{\mu}~=~x_{\mu}$.

The relations~(\ref{p1})-(\ref{s3}) are the commutations relations
for the set observables associated with a given observer. Another
observer's set satisfies the same commutations relations, pair of
sets being related by an element of the group SO(5,1). This means
that going from one observer to another, the observables mix up.
Given that, by construction, the sets includes space, time, and
matter observables, we can not disentangle space-time from matter.

The formalism of quantum mechanics tell us how to proceed. We must
construct the Hilbert space on which the observables act. A basis of
the Hilbert space is obtained selecting from that set, a complete
set of commuting observables. The algebra of the Lorentz group
SO(5,1) is of rank three and this means that the simultaneous
measurement of only \textit{three} observables completely determines
a particular state of the physical system. This is in marked
contrast with Poincare quantum mechanics where we need the four
components of the energy-momentum to label the physical states.
Moreover, here we may choose only one of the components to label
states, e.g., the energy $P^{0}$. We are then forced to choose, one
and only one of the components of the position three-vector, let's
say the component Z and together with it, the corresponding
component $J_{z}$ of the angular momentum.

Before turning to the consideration of the representation of the
homogeneous Lorentz group analogous to the one introduced by Dirac,
we may note in passing that the algebra of
observables~(\ref{p1})-(\ref{s3}) is invariant under the
automorphism defined by
\begin{eqnarray}
X_{\mu}\rightarrow\ P_{\mu},~~~
 J_{\mu \nu} \rightarrow J_{\mu \nu},~~P_{\mu}\rightarrow - X_{\mu}
 \nonumber \\
H \leftrightarrow \frac{1}{m_{pl}}.~~~~~~~~~~~~~~~  \label{aut}
\end{eqnarray}
To write a linear equation invariant under the group SO(5,1), we
follow Dirac's footsteps.  We construct a set of six irreducible
matrices~$\Gamma_{i}$ that satisfy the anticommutation relations
\begin{eqnarray}
\{\Gamma_{i},\Gamma_{j}\} &=&2 \eta_{ij}. \label{dir1}
\end{eqnarray}
The matrices  $\sigma_{ij}$
\begin{eqnarray}
\sigma_{ij} &=&-\frac{i}{2}[\Gamma_{i},\Gamma_{j}]  \label{dir2}
\end{eqnarray}
satisfy the commutation relations~(\ref{alg2}) and hence provide an
irreducible spinor representation~\cite{note4}.

We have at our disposal the differential operators
$L_{ij}$~(\ref{alg1}) and the matrices $\sigma_{ij}$ to work out a
linearized Casimir expression. An equation that satisfies the
requirements is
\begin{eqnarray}
H \sigma^{ij}L_{ij}\Psi(x)&=&\lambda \Psi(x), \label{dir3}
\end{eqnarray}
where $\lambda$ is a parameter with dimensions of energy.

This equation may be derived from the following Lagrangian
density\cite{note5}
\begin{eqnarray}
 \textit{L}(x)&=& - \bar{\Psi}(x)( H\sigma^{ij}L_{ij}-\lambda)\Psi(x)
\label{lag1}
\end{eqnarray}
with $\bar{\Psi}(x)=\Psi^{+}(x) \sigma^{60}$.

The lagrangian density~(\ref{lag1}) is invariant global gauge
transformations of the field $\Psi (x)$, $\Psi (x)
\rightarrow\exp(-i \alpha)\Psi (x)$. One can make it local, by
introducing a gauge field $A_{ij}(x)$ that transforms
$A_{ij}(x)\rightarrow A_{ij}(x)+i L_{ij}\alpha (x)$. The action
$\textbf{S}$ is then
\begin{eqnarray}
\textbf{S} &=&- \int d^{6}x  \bar{\Psi}(x)(
\sigma^{ij}D_{ij}-\lambda
 )\Psi(x) + \frac{1}{4 g}F^{2}  \label{act}
\end{eqnarray}
with $F_{ij~kl}=[D_{ij},D_{kl}]$ and  $D_{ij}=H L_{ij}+A_{ij}(x)$.
Note that the coupling constant g has dimension of length square.
The Euler-Lagrange equations are
\begin{eqnarray}
( H \sigma^{ij}L_{ij}-\lambda)\Psi(x)  &=&\sigma^{ij}A_{ij}(x)
 \Psi(x),  \label{eqn3} \\
 {H L_{ij}F^{ij}_{~~~kl}}(x)&=&g~\bar{\Psi}\sigma_{kl}\Psi(x). \label{eqn4}
\end{eqnarray}
The action~(\ref{act}) has in-built both an ultraviolet cutoff, the
Planck mass $m_{pl}$, as well as, an infrared one, the Hubble
constant H, that opens the door to the speculation that the
resulting quantum field theory is finite.

The author wishes to thank the Physics Department at Brandeis
University where most of this research was undertaken, and to
express his gratitude to J.~Peetermans and S.~Fraden for making it
possible. Support from the Central University of Venezuela is
acknowledged.


\begin{thebibliography}{99}
\bibitem{note1}
We use the word rotations for pseudo-rotations also.
\bibitem{note2}
We avoid taking the track that mantains the commutativity and leads
to the conformal group, but at the cost of requiring a nonlinear
realization for its physical interpretation.
\bibitem{note3}It has naturally the correct dimensions.
\bibitem{note4}These matrices are in principle $2^{3}$-dimensional.
\bibitem{note5}
It is straightforward to prove that if the lagrangian density
$\textit{L}$ is a functional of $\phi(x)$ and $L_{ij}\phi(x)$ only,
then the Euler-Lagrange field equations are $L_{ij}\frac{\partial
\textit{L}}{\partial \textit{L}_{ij}\phi(x)}-\frac{\partial
\textit{L}}{\partial \phi(x)}=0$.
\end{thebibliography}

\end{document}